\documentstyle[twoside,fleqn,espcrc2,epsfig]{article}
\newcommand{\beq}{\begin{equation}}
\newcommand{\eeq}{\end{equation}}
\newcommand\bea{\begin{eqnarray}}
\newcommand\eea{\end{eqnarray}}
\newcommand{\C}{{\cal C}}
\newcommand{\tr}{\mbox{Tr}}
\newcommand{\Past}{\mbox{\sl Past}}
\newcommand{\Future}{\mbox{\sl Future}}

\title{An insider's guide to quantum causal histories}

\author{Fotini Markopoulou\thanks{Email address: fotini@ic.ac.uk.}\\
The Blackett Laboratory, Imperial College,\\
            Prince Consort Road, South Kensington, London SW7 2BZ, U.K.\\
	    and\\
	    Center for Gravitational Physics and Geometry, Department of 
	    Physics,\\
	    The Pennsylvania State University, University Park, PA 16801, USA.}
       
\begin{document}

\begin{abstract}
A review is given of recent work aimed at constructing a quantum
theory of cosmology in which all observables refer to information
measurable by observers inside the universe.  At the classical level 
the algebra of observables should be modified to take
into account the fact that observers can only give truth values
to observables that have to do with their backwards light cone.
The resulting algebra is a Heyting rather than a Boolean algebra.
The complement is non-trivial and contains information about
horizons and topology change.  Representation of such observables
quantum mechanically requires a many-Hilbert space formalism,
in which different observers make measurements in different Hilbert
spaces.  I describe such a formalism, called ``quantum
causal histories'';  examples include causally
evolving spin networks and quantum computers.
\end{abstract}

\maketitle

\section{Introduction}

A quantum theory of gravity is expected to also be a satisfactory theory 
of quantum cosmology.  In turn, a quantum theory of cosmology would 
only be acceptable if it admits a description fully from within 
the universe itself.
We may translate this into the requirement that, 
in a satisfactory quantum theory of gravity, the physical observables 
must refer to observations made inside the universe.  

This has immediate consequences.  Consider, for example, the familiar 
construction of the 3-geometry wavefunction that is used in the 
Wheeler-DeWitt equation.  This describes the quantum state of a 
spatial slice.  However, in a causal spacetime, only very special
observers, such as an observer at the final singularity,
can have access to an entire slice.  Since there are no observers
outside the universe, this wavefunction is not an observable quantity.

In \cite{Mark98c}, I argued that what is 
required to construct a cosmological theory is {\em internal observables}, 
corresponding to observations made inside the universe.  Internal 
observations contain only partial information about the universe, 
that which is in the causal past of an observer at the corresponding 
spacetime region.   It then becomes desirable to set up a framework 
for quantum gravity in which all physical observables are internal.
To see how to do this we begin by understanding the effect the
requirement that all observables are internal has on the
structure of observables in classical general relativity.

\section{Classical Internal observables and their algebra}

Let us start by considering only the causal structure of the 
universe.   To proceed it is convenient to approximate the
causal structure of spacetime by picking out a discrete set of
events. This gives us an  approximate description of 
the causal structure of the
spacetime in terms of a 
causal set \cite{Bomb87}.   This is s set of events $p,q,r,\ldots$, ordered by a 
causal preceding relation $p\leq q$, which is transitive ($p\leq q$ 
and $q\leq r$ imply that $p\leq r$), is locally finite (given $p\leq 
q$ the intersection of the past of $q$ and the future of $p$ contains 
a finite number of events), and has no closed timelike 
loops (if $p\leq q$ and $q\leq p$, then $p=q$). 

In such a causal set universe, the internal observables are {\rm 
functors} from the causal set to the category of sets\footnote{
A category is a collection of objects, and arrows between the objects.  
There should be a unit, and the composition of arrows should be 
associative.  Thus, a causal set is a category (partial order) whose 
objects are the events, and arrows are the causal relations.  ${\bf 
Set}$ is the category whose objects are sets and whose arrows are maps 
between sets.  A functor can be thought of as a ``function'' from one 
category to another that turns the objects of the first into objects 
of the second, while preserving the properties of the arrows 
of the first into the second.}.  
Details can be found in \cite{Mark98c}, here we will simply discuss examples.  
An prime example of an internal observable then is the one describing 
causal past.  This is the functor 
\beq
\Past :\C\longrightarrow {\bf Set},
\eeq
that outputs events that have occurred.  It
has components at each event $p$ which are the causal past of $p$:
$\Past(p)=\{r\in\C :r\leq p\}$.  Further, the functor contains 
not only all these sets, but the maps between them:  
$\Past(p)\subseteq \Past(q)$, whenever $p\leq q$.

The internal observable $\Past$ can be thought of as a 
varying set, having components at each event which are sets, all tied 
together by inclusion functions.  Thus, the causal structure of the 
universe is built into the observable. 
These are fundamentally different than standard set-like observables.  
This can be seen by considering the algebra of internal observables.  

The algebra of standard (fixed time) observables is obtained by 
is called a 
Heyting algebra.  As a result a theory with internal observables 
is fundamentally different that a theory describing 
a system external to the observers.  It has a different logical 
structure.  Just as the Boolean algebra obeyed by set-like 
observables means that physical propositions obey boolean logic, 
physical propositions in a theory with internal observables obey 
{\em intuitionistic} logic.  Its characteristic feature is that, for some 
statement $a$, $\neg\neg a \neq a$.  
Thus,  the requirement that physical observables
must refer to measurements made by observers in the universe has
as a consequence the fact that the very logic that observers
use to describe what they see must be modified.  It should take into account
the fact that any single observer is only able to know a subset
of the true facts about their universe.

An immediate application of the Heyting algebra is in coding the 
causal topology.  It is easy to check that in a universe with an 
initial event, which causally preceeds all the others, 
$\Future(p)=\emptyset$, for 
all $p\in\C$.  In a universe with a final event, in the future of all 
others, $\Past(p)=\emptyset$ for all $p\in\C$.  And for a universe 
with both, both internal observables have all their components empty.  
As a result, as described in \cite{Mark98c}, the existence of horizons
and topology change can be deduced from the Heyting complements.

\section{The framework of Quantum Causal Histories}

A quantum cosmological theory should involve only internal observables;
thus it should have an algebra of observables
whose  classical limit is a Heyting algebra of the kind
we just discussed.  But we know that the $\hbar \rightarrow 0$
limit of the projection operators on an ordinary Hilbert space is
a Boolean algebra.  Therefore we need to look for a formulation of
quantum cosmology that is not based on the usual single Hilbert space
formalism.  One possible way to proceed 
is to ``quantize'' the causal structure by  attaching Hilbert spaces to the 
events of a causal set.   These can be thought of as elementary 
Planck-scale systems that interact and evolve by rules that give rise 
to a discrete causal history.   An example of such a theory is the 
causal evolution of spin networks \cite{Mark98a}, as we will see in the 
next section.  But let us first give a brief discussion of the basic 
features of quantum causal histories (or QCH for short).  

\subsection{Hilbert spaces on the events}

Consider a causal set $\C$.  This is a ``spacetime'' graph, with 
nodes which represent events, and directed edges coding the causal 
ordering of the nodes.    Let us interpret the events as elementary 
quantum mechanical systems, which are to be encountered at Planck 
scale.  Thus, we may attach a Hilbert space to each node of the 
causal set graph, representing the elementary system that we encounter 
at that node.  Since we are building a theory with a fundamental 
discreteness in the causal relations between these systems, thus 
assuming there is a lowest scale, it is reasonable to expect that 
these Hilbert spaces are finite-dimensional.  We have, therefore, 
built a causal network of finite-dimensional Hilbert spaces.  

By the standard rules of quantum mechanics, the combined state space 
of a set of events that are acausal to each other is the tensor 
product of the individual Hilbert spaces.  

When may we expect that there is unitary evolution in this quantum 
causal history?  It is not difficult to check that this is only 
possible between acausal sets of events $a$ and $b$ that form a 
complete pair $a\leq b$ , that is, every event in $b$ is to the future 
of some event in $a$ and every event in $a$ is to the past of some event 
in $b$, as shown ($H_{a}=H_{1}\otimes H_{2}$ and $H_{b}=H_{3}\otimes 
H_{4}\otimes H_{5}$):
\beq
	\begin{array}{c}\mbox{\epsfig{file=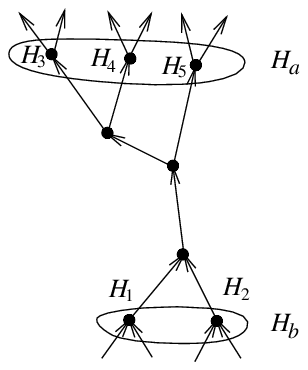}}\end{array}
\nonumber
\eeq

Any information that reaches $b$ has come 
through $a$, and there is no event in the future of $a$ which is not 
related to $b$.  Information is therefore conserved from $a$ to $b$ 
when a unitary evolution map relates the two Hilbert spaces.  

It should be emphasized that this is {\em local} unitary evolution, in 
the sense that the complete pair $a$ and $b$, in general, are 
localised spacelike regions in the universe.  In the special case
when the causal set admits a global foliation into a set of 
antichains (maximal sets of events in the causal set that are 
all acausal to each other), there is a linear sequence of 
unitary evolution operators  (this may be compared to 
quantum field theory on a globally hyperbolic spacetime).

\subsection{Hilbert spaces on the edges}

Closer inspection of the above model reveals that the unitary 
evolution operators do not, in general, respect local causality.  As 
an example, consider the following configuration:
\beq
	\begin{array}{c}\mbox{\epsfig{file=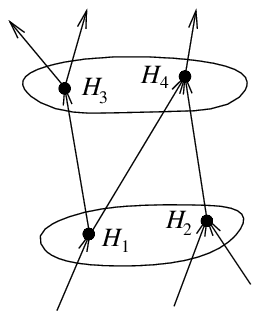}}\end{array}
\nonumber
\eeq
with evolution operator $E:H_{1}\otimes H_{2}\rightarrow H_{3}\otimes
H_{4}$.  Next, select some state $|\psi_{2}\rangle\in H_{2}$.  This 
can be extended to $|\psi_{1}\rangle \otimes  |\psi_{2}\rangle$ in 
$H_{1}\otimes H_{2}$ and then evolved by $E$ to some $|\psi'\rangle
\in H_{3}\otimes H_{4}$.  Finally, $H_{4}$ can be traced out to leave 
a density matrix $\rho_{3}=\tr_{H_{4}}|\psi'\rangle\langle\psi'|$ in 
$H_{3}$.  That is, system 3 ``knows about'' $|\psi_{2}\rangle$, even 
though there is no causal link from 2 to 3.  This is a violation of 
the causal relations of the underlying causal set.  

There is a straightforward solution to this problem.  Instead of 
attaching the Hilbert spaces to the events of the causal set, let us 
attach them to the {\em causal relations}.  We again take tensor 
products of Hilbert spaces on edges that are acausal to each other.  
Any unitary operator in a history where the Hilbert spaces are on the 
edges can be decomposed to a product of unitary operators that live on 
the nodes of the causal set, going from the composite Hilbert space on 
the incoming edges to that node to the composite Hilbert space on the 
outgoing ones, all of which respect the causal structure of $\C$.
Therefore, in a QCH with the Hilbert spaces on the 
causal relations and the operators on the events, the quantum evolution 
strictly respects the underlying causal set.  

We may also note that promoting the nodes of the causal set to 
evolution operators is consistent with the intuition that an event in 
the causal set denotes change, and so is most naturally represented by 
an operator. In addition, since only spacelike separated Hilbert 
spaces are tensored 
together, there is no single Hilbert space, or wavefunction, for the 
entire universe.

\section{Examples of quantum causal histories}

We will now give specific examples of QCH models.  
To do so, we need to identify the Hilbert spaces and the complete 
pairs that are related by the unitary evolution operators.  
The first of our two examples is causal spin network evolution, a model 
of quantum spatial geometry evolving causally.  The second example 
considers identical individual Hilbert spaces which are 
two-dimensional, and the resulting QCH is a quantum 
computer.  

\subsection{Causal evolution of spin networks}

Spin networks were originally defined by Penrose as trivalent graphs 
with their edges labelled by representations of $SU(2)$ 
\cite{Penrose}.  From such abstract labelled graphs, Penrose was able to 
recover directions (angles) in 3-dimensional Euclidean space.  Later, 
in loop quantum gravity, spin networks were shown to be the basis 
states for the spatial quantum geometry states \cite{loopspinnets}.  
The quantum area and volume operators, expanded in the spin network 
basis were shown to have a discrete spectrum, with their eigenvalues 
depending on the labels of the spin network present in the region of 
space whose area, or volume, is being measured.  

In \cite{Mark97,Mark98a}, spin network graphs were used as model of 
quantum spatial geometry evolving causally.  This means that the nodes 
of the spin network graph are the events in a causal set.  

A causal spin network history is a quantum causal history.  To see 
this, we need to observe that it has the following features.  The 
Hilbert spaces are the spaces of intertwiners.  An intertwiner 
labels a node of a spin network.  It is a map from the tensor product 
of the representations labelling the edges incoming to that node to the 
tensor product of the  outgoing edges.  The possible intertwiners for 
a node in the spin network form a vector space, the so-called ``space of 
intertwiners''\footnote{For a trivalent node of an $SU(2)$ spin 
network, the intertwiner is unique.  For a four-valent node, the 
intertwiner space is finite-dimensional.  For higher valence, 
continuous parameters enter.}.   

The intertwiner spaces of spacelike separated nodes in the causal 
history are to be tensored together, with the representations on any 
connecting edges summed over:
$V_{ijkmno}=\sum_{l} V_{ijkl}\otimes V_{lmno}$ when $l$ labels the 
shared edge.

In \cite{Mark97}, it was shown that, if the spin networks were restricted to 
be of valence $n$, a small set of generating evolution operators can 
be identified.  These are the 1-skeletons of the $n$-dimensional 
Pachner moves for piecewise linear triangulations \cite{Pachner}.  
For example, in 2+1, the set of elementary moves is shown below; in 
each pair the top and bottom configurations are exchanged.
\beq
	\begin{array}{c}\mbox{\epsfig{file=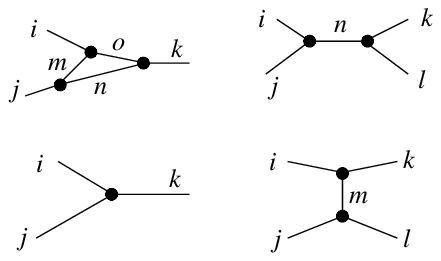}}\end{array}
\nonumber
\eeq
Given an initial 3-valent spin network, to be thought of as modeling a 
quantum ``spatial slice'', the causal history is built by repeated 
application of the above moves.  They change the spin network 
locally, thus producing a discrete analogue of multifingered time 
evolution.   Thus, the amplitude to go from an given initial spin 
network $\Gamma_{1}$ to a final one $\Gamma_{2}$ can be expressed as 
the product of the amplitudes for the Pachner moves that occur in a 
spacetime history extrapolating between the two spin networks, summed over  
the possible extrapolating histories:
\beq
A_{\Gamma_{1}\rightarrow\Gamma_{2}}=
\sum_{\mbox{\footnotesize histories }\Gamma_{1}
           \rightarrow\Gamma_{2}}
	   \prod_{
	       \begin{array}{c}
	       \mbox{\footnotesize moves in a} \\
	       \mbox{\footnotesize given history}
	   \end{array}}
	   A_{\mbox{\footnotesize move}}
	   .
\eeq
Explicit expressions for the amplitudes $A_{\mbox{\footnotesize move}}$ 
for the elementary moves
have so far only been given for a simple causal model in \cite{Borr97}.  

By construction, the Pachner moves are always moves between complete 
pairs (they are homs of the spin network graph).  Thus, they can be 
consistently promoted to unitary operators.  

This completes the identification of causal spin network histories as 
a QCH model.  The individual Hilbert spaces are 
the intertwiner spaces, which are to be tensored when they are 
spacelike separated in the history.  The local unitary operators are 
the Pachner moves.  

Having performed these identifications, we may note that
several more models of the same type have been explored in the 
literature.  They are graphs evolving under the above causal moves, 
but with different sets of labels.  Trivalent graphs labelled by 
ratios of integers give rise to Sen's string networks 
\cite{Mark98b,Sen}.  Using $q$-deformed spin networks, that is, spin 
networks labelled by representations of $SU_{q}(2)$ have a finite list 
of labels that may appear on an edge \cite{qspinnets}.  In fact, spin 
networks can be constructed  that are labelled by representations of 
any compact group \cite{Mark98a}, as well as supersymmetry 
\cite{susy}, and all give rise to quantum causal histories when evolved 
causally.

\subsection{Quantum computers}

Possibly the simplest choice of individual Hilbert spaces in a 
QCH is to require that they all are 
2-dimensional: ${\bf C}^{2}$.  Having done so, it is unavoidable to 
note that these spaces are qubits and the history is a (very large) 
quantum computer!  (For related work, see \cite{PZ}).
A choice of local unitary operators is a choice of quantum gates in a 
quantum computer.  The set of properties of the underlying causal set 
is identical to the computer's circuit.  

Given how hard the task of finding explicit expressions for suitable 
QCH evolution operators,  this model provides the opportunity to
use the quantum gates used in quantum computing to model 
quantum spacetime evolution.  It is possible that 
there is a relationship between the conditions required for a quantum 
computer to run for a long time and a quantum spacetime to have a 
classical limit.  

\section{What a quantum causal history looks like from the Inside}

We may now briefly return to internal observables and outline how we 
expect they will appear in QCH.  Consider a QCH with Hilbert spaces 
labelling the causal relations, and let us interpret them in a way 
that will help us set up internal observables.  

Given some event $p$ in the causal set, let $q$ and $r$ be two events 
in the future of $p$.  In the QCH on this causal set, there are two 
Hilbert spaces for the two causal relations, $H_{pq}$ and $H_{pr}$ 
respectively.  We will interpret the first as ``the state space of 
$p$ as seen by $q$'', and the second as ``the state space of $p$ as 
seen by $r$''.   The relation between the two should depend on the 
causal relation between $q$ and $r$.   Thus, if $S(p)$ is the set of 
causal relations that start at $p$,  there is a Hilbert space for 
every element of $S(p)$, describing how the an observer at the end of 
that particular causal relation sees $p$. 

We may then define a {\em generalised} Hilbert space for $p$ to be a 
functor $H_{p}:\C\rightarrow \mbox{\sl Hilb}$, which has as its 
elements the individual ``viewpoint'' Hilbert spaces, linked together 
by consistency maps that transform from the viewpoint of one 
observer to the viewpoint of another.  A standard (not 
observer-dependent) description is recovered when these consistency 
maps are identities.   Then $H_{p}$ becomes a 
standard state space for $p$.

It is on such generalised Hilbert spaces that the quantum internal 
observables are expected to act. A quantum internal observable
should be a generalized operator, by which is meant an operator
on each of the components of a generalized Hilbert space, related
by the consistency maps.  Details of this construction will
appear elsewhere.

\section{Conclusions: General Relativity as the low energy limit of 
quantum gravity}

In the above examples we have seen that the requirement that
all observables are internal has non-trivial consequences for the
structure of both classical and quantum cosmological theories.
One should not forget, however, that any Planck scale quantum
cosmological theory will have to have general relativity as its
low energy limit.  We have not discussed this aspect of quantum
gravity here, but progress on methods to obtain the low energy
limit is needed in order to bring the developments described here
to a conclusion.  Work is in progress currently on methods to
coarse grain and renormalize quantum causal histories, which
will be reported elsewhere.
 
\section*{Acknowledgments}
I am grateful to John Baez, Louis Crane, Sameer Gupta, Eli Hawkins, 
Louis Kauffman, Renate Loll, Carlo Rovelli and Lee Smolin
for discussions.   I especially want to thank Chris Isham for 
explaining to me the use of functorial methods in quantum theory in 
\cite{CJI} and many discussions.  I would also like to thank Abhay Ashtekar 
for hospitality at the Center for Gravitational Physics and Geometry 
during the course of this work.   

This work was supported by NSF grant PHY/9423950 and PHY/9514240, 
and a gift from the Jesse Phillips foundation.

\end{document}